\documentclass[useAMS,usenatbib]{mn2e}
\usepackage{graphicx}
\usepackage{fixltx2e}
\title[Fast X-ray transients in the Galactic bulge with 
the {\it Rossi X-ray Timing Explorer}]{Fast X-ray transients toward the Galactic bulge with 
the {\it Rossi X-ray Timing Explorer}}

\author[D. M. Smith, C. B. Markwardt, J. H. Swank, and I. Negueruela]
{D. M. Smith$^{1,2}$\thanks{E-mail: dsmith8@ucsc.edu}, 
C. B. Markwardt$^3$, J. H. Swank$^3$, and I. Negueruela$^4$ \\
$^{1}$Physics Department and Santa Cruz Institute for Particle Physics, University of California, 
Santa Cruz,\\
 1156 High St., Santa Cruz, CA 95064, USA\\
$^{2}$Space Sciences Laboratory, University of California, Berkeley, Berkeley, CA, USA\\
$^{3}$NASA's Goddard Space Flight Centre, Greenbelt, MD, USA\\
$^{4}$Departamento de F\'{\i}sica, Ingenier\'{\i}a de Sistemas y Teor\'{\i}a de la Se\~{n}al, Universidad \\
de Alicante, Alicante, Spain }

\begin{document}

\date{Accepted ??. Received ??; in original form ??}

\pagerange{\pageref{firstpage}--\pageref{lastpage}} \pubyear{2010}

\maketitle

\label{firstpage}

\begin{abstract}

In X-ray binaries, rapid variability in X-ray flux of greater than an
order of magnitude on time-scales of a day or less appears to be a
signature of wind accretion from a supergiant companion.  When the
variability takes the form of rare, brief, bright outbursts with only
faint emission between them, the systems are called Supergiant Fast
X-ray Transients (SFXTs).  We present data from twice-weekly scans of
the Galactic bulge by the {\it Rossi X-ray Timing Explorer (RXTE)} that
allow us to compare the behaviour of known SFXTs and possible SFXT
candidates with the persistently bright supergiant X-ray binary
4U~1700--377.  We independently confirm the orbital periods reported by other groups
for SFXTs SAX~J1818.6--1703 and IGR~J17544--2619.  The new data do
not independently reproduce the orbital period reported for
XTE~J1739--302, but slightly improve the significance of the original
result when the data are combined.
The bulge source XTE~J1743--363 
shows a combination of fast variability and a long-term decline in
activity, the latter behaviour not being characteristic of 
supergiant X-ray binaries.  A far-red spectrum of the 
companion suggests that it
is a symbiotic neutron star binary rather than a high-mass binary,
and the reddest known of this class: the spectral type is approximately
M8\,III.

\end{abstract}


\section{Introduction}

Blue supergiant stars were first identified with fast X-ray transients
discovered by {\it ASCA} (AX~1845.0--0433, \citet{yamauchi95,coe96})
and later by {\it RXTE} (XTE~J1739--302, \citet{smith98,smith03}). As more
began to be discovered with {\it INTEGRAL}, they were recognised as a
class \citep{smith04,zand05,negueruela06,sguera05}
and named Supergiant Fast X-ray Transients (SFXTs) 
\citep{negueruela06}.

SFXTs are thought to be a subclass of X-ray binary in which a neutron
star accretes the wind of a supergiant companion.  Some binaries with
these two components have been known since the 1970s as very bright
and variable X-ray sources (e.g., Vela~X-1, 4U~1700--377), but the
SFXTs spend most of their time at low luminosities of $10^{32}$ to $10^{34}$
erg s$^{-1}$, with only very brief excursions to outburst luminosities
of up to a few times $10^{36}$ erg s$^{-1}$.  These outbursts
generally last for several hours, with subpeaks lasting for minutes
(e.g. \citet{sidoli09,rampy09}).

The cause of this behaviour is not known.  Clumps in the wind of the
companion can explain rapid variability within outbursts
\citep{zand05,walter07,negueruela08a}, but are less successful at
explaining long periods of quiescence.  In some cases a highly
eccentric or wide orbit may be responsible \citep{sidoli07}, but some
SFXTs have been shown to have orbits as short as the persistently
bright sources (see Table~1).  Rejection of the wind material by the
magnetic propeller effect \citep*{grebenev07, bozzo08a,li11} has been
suggested.  Depending on the particular systems studied and the assumptions
made about the accretion physics, consideration of the propeller effect
has led to both high lower limits ($>10^{13}$~G) \citep{bozzo08a} and
low upper limits ($>10^{10}$~G) \citep{li11} on the magnetic field.
Unexpected structure in the companion wind itself,
e.g. very rare clumps with a very high density contrast,
has not yet been ruled out.

Some sources have been referred to as intermediate between SFXTs and
the familiar persistent supergiant/neutron star binaries, but this
category has never been universally defined.  The most
common criterion for ``intermediateness'' mentioned in the
literature is the dynamic range between the highest and lowest
luminosities observed.  ``Intermediate'' sources by this criterion
have been defined as either those with a dynamic range of $<$100 but
$>$20 or so \citep{walter07,clark10} or those with a range of
100--1000 \citep{sguera11,romano12}.  
The maximum dynamic range of any source --
both the SFXTs and the classical wind-accreting HMXBs such as
Vel~X--1 -- can only increase with the number of observations made.
Thus it is not surprising that the most recent of these references
give the higher range for intermediate systems, since the known
dynamic ranges of all the objects have increased in recent years.
\citet{chaty10} adds two additional criteria to the definition of
intermediateness: a higher average luminosity than the canonical
SFXTs, and a longer average duration for individual flares.
\citet{romano12} add the criterion that the lowest luminosity
frequently observed should be close to $10^{35}$ erg~s$^{-1}$ for intermediate
systems, and 1 to 2 orders of magnitude lower for true SFXTs.
\citet{romano11} have begun to make the SFXT vs. intermediate 
classification systematic for the first time by
interpreting histograms of the luminosity of three sources,
XTE~J1739--302, IGR~J17544--2619, and IGR~J16479--4514 made from
monitoring data from the sensitive {\it Swift} X-Ray Telescope (XRT).
The difference between true SFXT and intermediate systems in that
work is based on the fraction of time spent at the lowest luminosities.

\section{Data and analysis}

We present data from the Galactic bulge scan programme \citep{swank01}
of the Proportional Counter Array (PCA) instrument on the {\it Rossi
X-ray Timing Explorer (RXTE)} \citep{jahoda06}.  PCA is a large-area
X-ray detector with a collimated but non-imaging field of view of
1$^{\rm{o}}$ full width at half maximum.  Twice a week, PCA scans the
inner Galaxy in a raster pattern, recording the instantaneous
2--10~keV count rate of a large catalogue of sources.  The bulge scan
data are publicly available.
\footnote{http://lheawww.gsfc.nasa.gov/users/craigm/galscan/main.html}

\begin{figure*}
\includegraphics[width=0.9\textwidth]{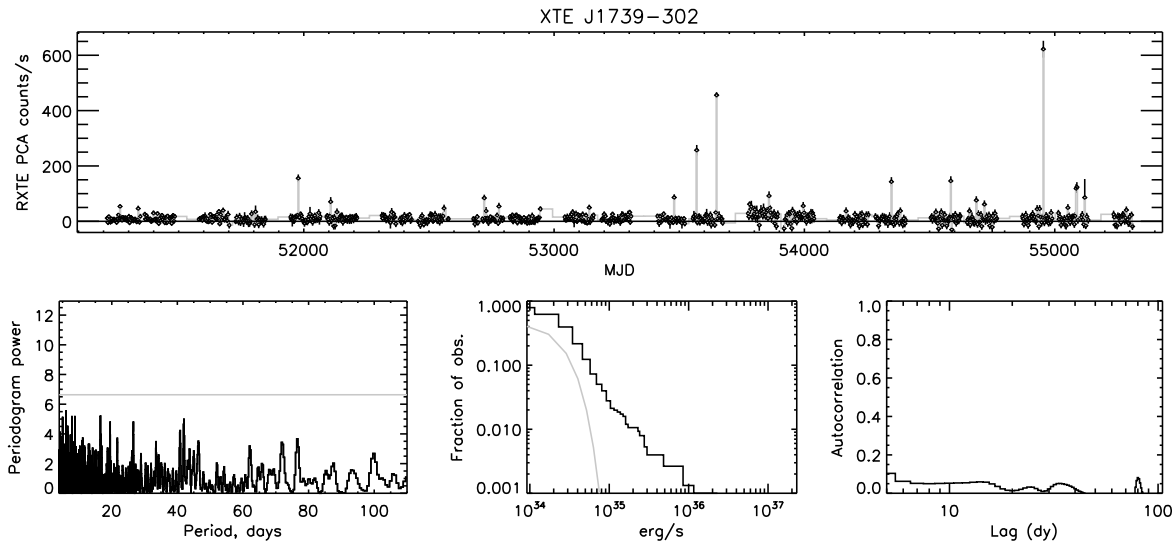}
\caption{X-ray time series analysis for XTE~J1739--302.  {\bf Top} (panel a):
X-ray count rate as a function of time from the bulge scan programme.
The rate shown is the equivalent rate for all 5 detector units (PCUs)
summed over the 2--10~keV band, even though fewer detectors were often
operating, particularly later in the mission.  {\bf Bottom left} (panel b): Lomb
normalised periodogram of the data in panel a.  The grey line represents
the average maximum value in each period bin in 10$^4$ randomised trials.
{\bf Bottom middle} (panel c):  histogram of the fraction of samples above a given
luminosity as a function of luminosity, using the first-listed distances from Table~1
(see text).  {\bf Bottom right} (panel d): autocorrelation function of the data
in panel a.}
\end{figure*}

\begin{figure*}
\includegraphics[width=0.9\textwidth]{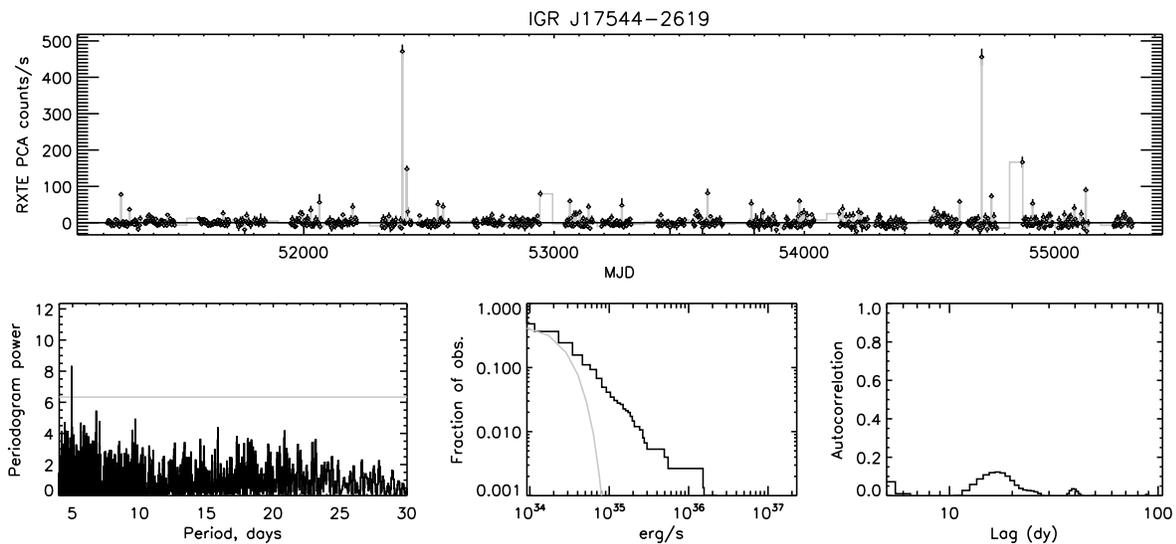}
\caption{Data products as described in Fig.~1, for
IGR~J17544--2619.}
\end{figure*}

\begin{figure*}
\includegraphics[width=0.9\textwidth]{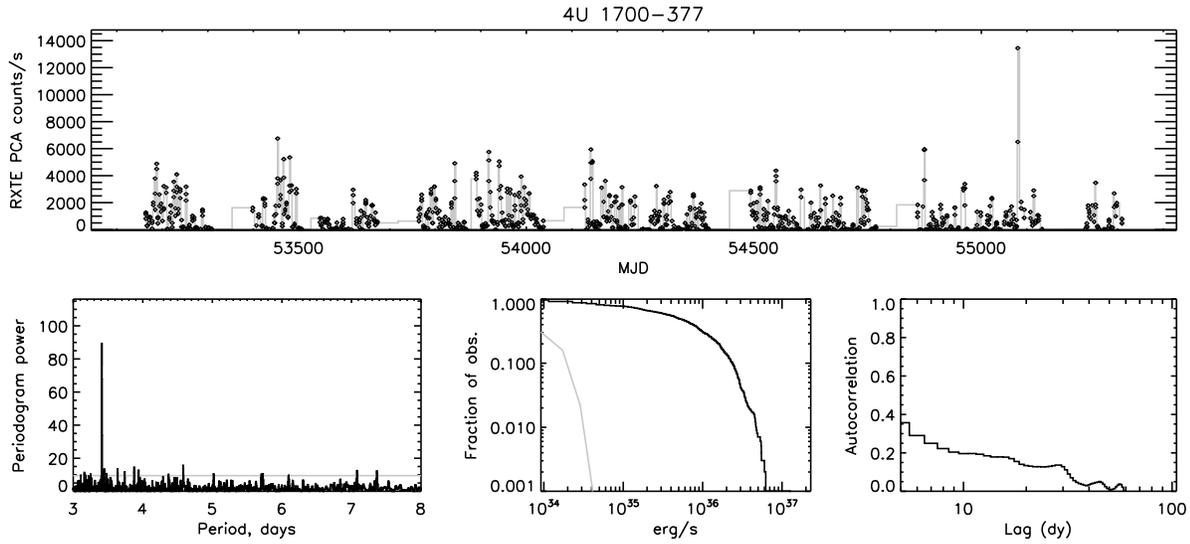}
\caption{Data products as described in Fig.~1, for
4U~1700--377.}
\end{figure*}

\begin{figure*}
\includegraphics[width=0.9\textwidth]{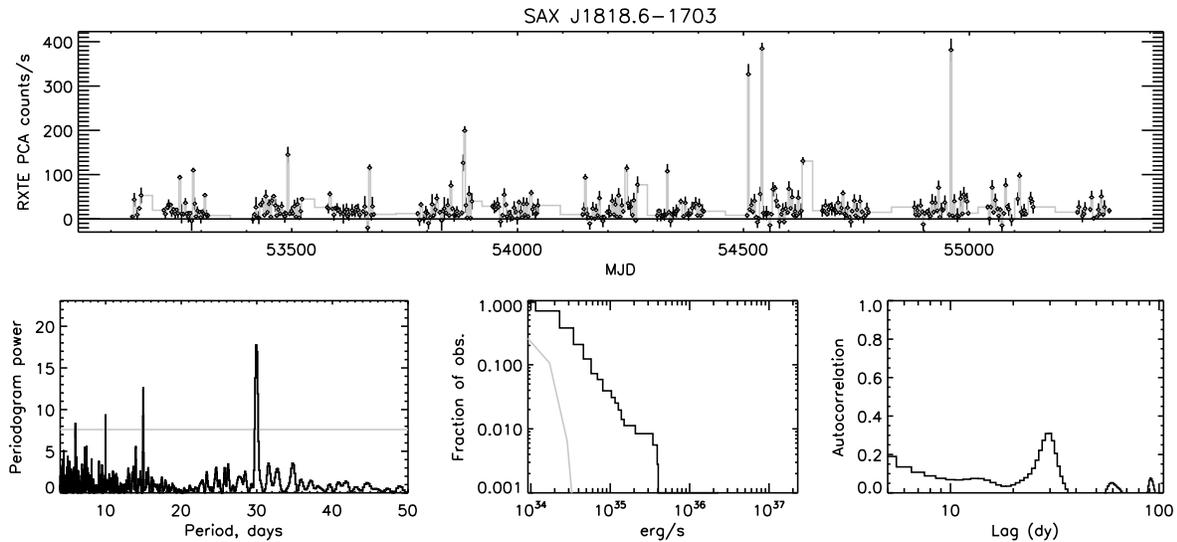}
\caption{Data products as described in Fig.~1, for
SAX~J1818.6--1703.}
\end{figure*}

\begin{figure*}
\includegraphics[width=0.9\textwidth]{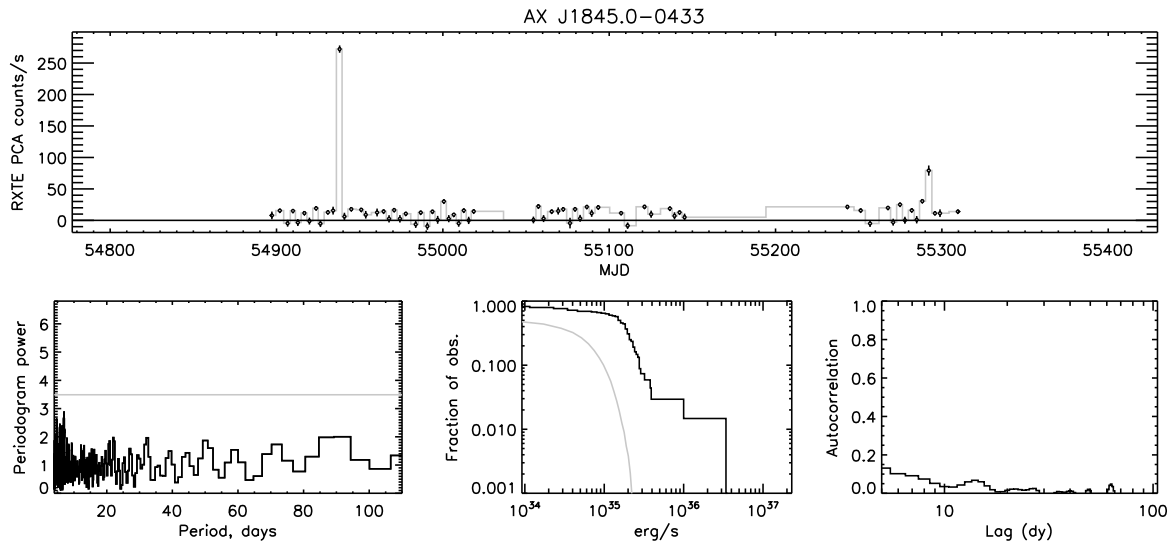}
\caption{Data products as described in Fig.~1, for
AX~J1845.0--0433.}
\end{figure*}

\begin{figure*}
\includegraphics[width=0.9\textwidth]{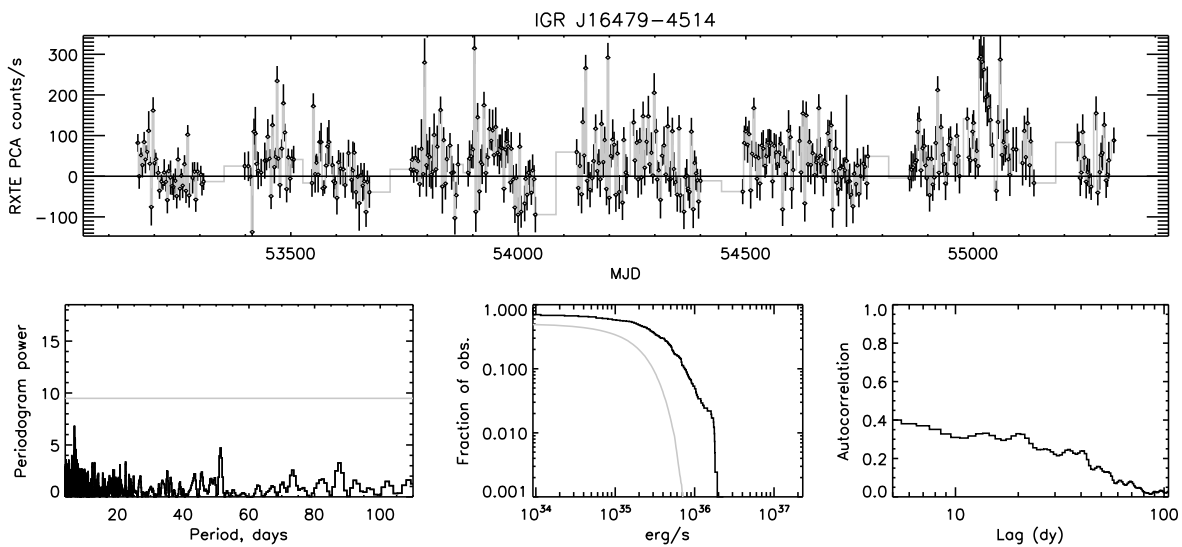}
\caption{Data products as described in Fig.~1, for
IGR~J16479--4514.}
\end{figure*}

\begin{figure*}
\includegraphics[width=0.9\textwidth]{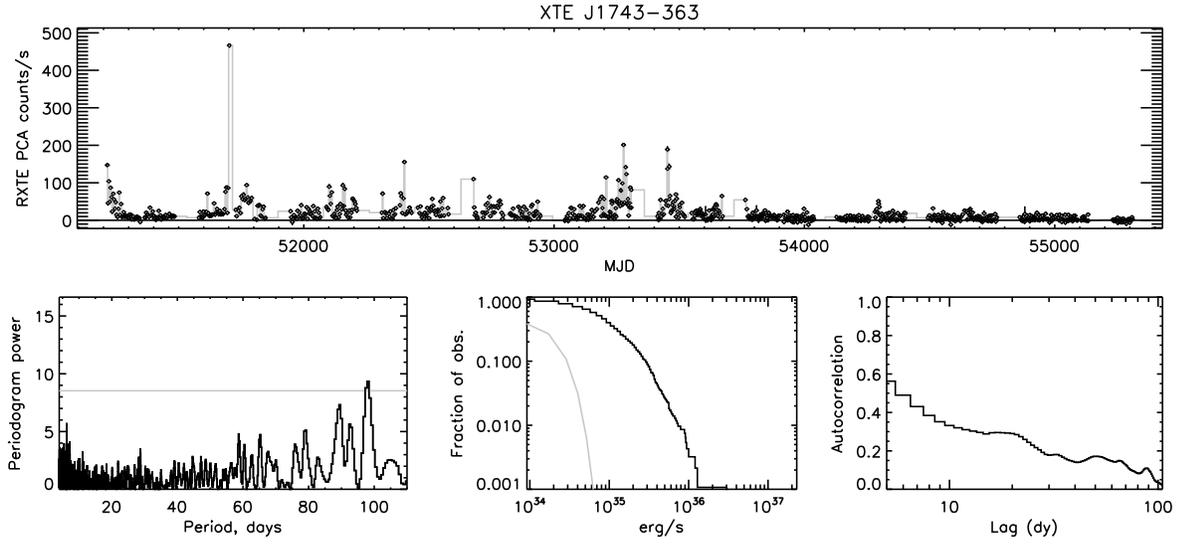}
\caption{Data products as described in Fig.~1, for
XTE~J1743--363.}
\end{figure*}

\begin{figure*}
\includegraphics[width=0.9\textwidth]{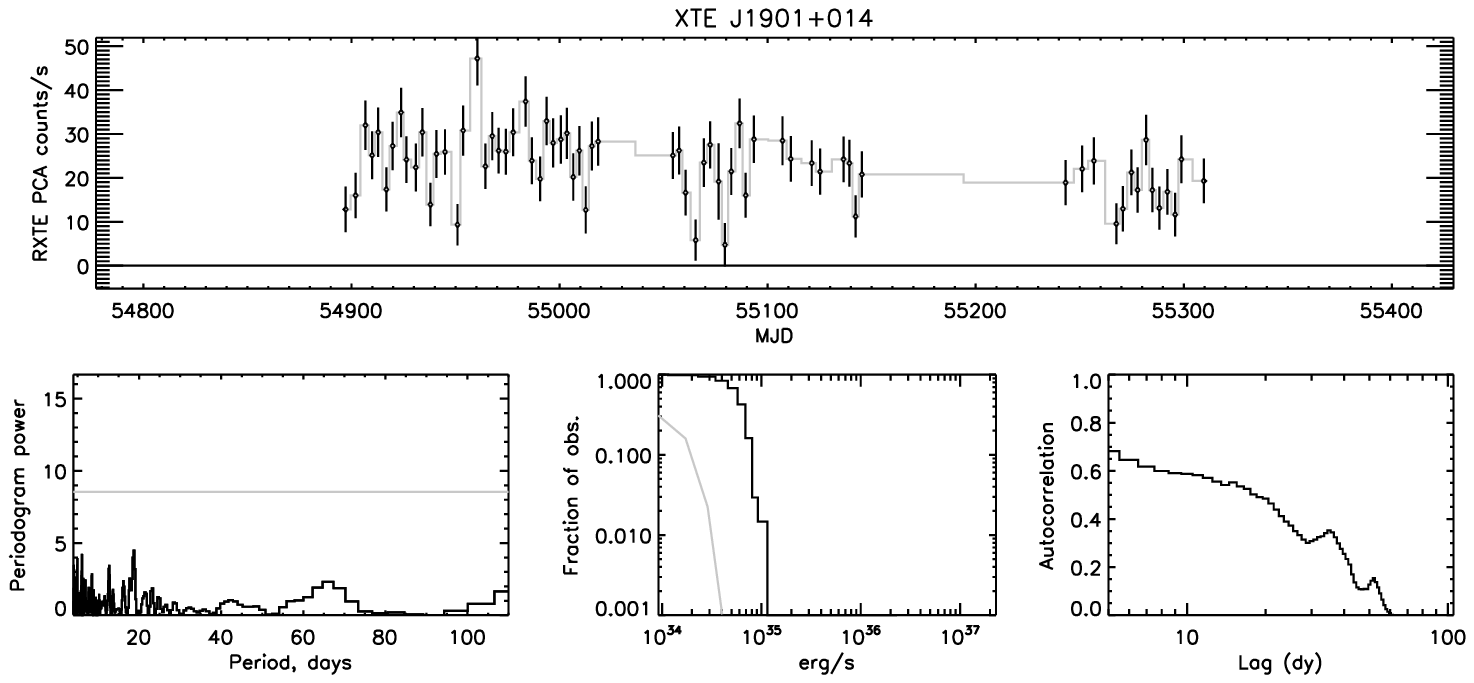}
\caption{Data products as described in Fig.~1, for
XTE~J1901+014.}
\end{figure*}

In this paper we show data from the two 'prototype' SFXTs
XTE~J1739--302 and IGR~J17544--2619 (Fig.~1 \& 2), the persistently
bright wind accretor 4U~1700--377 (Fig.~3), three supergiant
binaries that are not as well studied and have been considered either
SFXTs or 'intermediate' systems (SAX~J1818.6--1703, AX~J1845.0--0433,
and IGR~J16479--4514, Fig.~4, 5, and 6, respectively), and finally
two fast transients whose companion type and nature are not known:
XTE~J1743--363 and XTE~J1901+014 (Fig.~7 and 8).  For each source
we present four complementary views of the data: the 2--10~keV
light curve from up to 12 years of bulge scans, a Lomb-normalised
periodogram used to search for orbital periodicity, an integral
histogram of the luminosities, and the autocorrelation function of the
light curve.

The light curves can be considered reliable at relatively high count
rates, but near zero counts the results are complicated by the
systematic limitations of the technique of fitting the count-rate time
profiles of the raster scans.  For example, systematic uncertainties
in the profile of the Galactic 'diffuse' emission (much of which may
be an ensemble of many faint cataclysmic variables and other stars -- see
\citet{revnivtsev06}) translate to uncertainties in the lowest source count
rate.  This problem is particularly severe for AX~J1845.0--0433
(Fig.~5a), where the bifurcated appearance of the baseline values is
caused by the alternation of scans parallel and perpendicular to the
Galactic plane. The very large error bars for IGR~J16479--4514
(Fig.~6a) are due to the proximity of the bright low-mass X-ray
binary GX~340+0.  We therefore cannot draw conclusions about the
fluxes of the latter two objects outside of outburst.  The counts 
shown are normalised to be the expected 2--10~keV count rate if all 
five detectors making up the PCA were turned on.  In practice, from one to
five detectors are on at any given time, and this will be reflected
in varying error bars from point to point.  In these units, the
Crab would be on the order of 10000 counts s$^{-1}$. 

The second panels (b) of Fig.~1--8 show the Lomb-normalised periodogram of
the data, used to search for orbital periods.  The gray horizontal
line represents the largest expected value for each period in $10^4$
randomised trials (the real data values were randomly reassigned to the
real data acquisition times in each trial and the periodogram taken).
The output of the simulations has a slightly different maximum value
for each value of period, but there is no significant dependence of
the sensitivity on period; the variation looks like noise and does not
maintain repeatable features when the initial random number seed 
of the full set of randomised trials is changed.
Thus we replace the noisy sensitivity level of the simulations with a
constant value equal to the average of all the period bins to give the
line shown, which is the 99.99 per cent significance threshold for the case
of a known period.  

The third panel (c) of each figure shows an integral luminosity
histogram (the fraction of measurements above each flux value), with
the fluxes calculated from the count rates using the PCA instrument
response and the distances shown in Table~1 (where more than
one distance is cited, the first is used).
Systems with no good distance estimate (XTE~J1901+014 and XTE~J1743--363)
have been placed at 3~kpc, a distance typical of
the estimates for sources that have them.  If the nature of
these two objects is completely different from that of
the supergiant binaries, their distances could be quite different as
well.  For example, if either is a cataclysmic variable,
the distance and luminosity
could be much less.  Observation of most of these sources in the
Galactic bulge is probably due to the region being well observed
rather than any physical association with the bulge populations; note
that the distances derived for the sources with known companions are
all well in the foreground of the Galactic Centre (Table~1).  In these
panels, the gray lines shown for comparison are the integral of a
normal distribution centred at zero whose standard deviation is equal to the average
error bar for the lowest 20 per cent of flux observations from each source.
This represents the expected distribution if there were no real flux
from the source but only random noise.

Finally, the last panel (d) of each figure gives the autocorrelation
function of the data for each system.  To generate this curve, we
interpolated the time history of each system on to a 1~dy uniform grid,
setting all the values to zero during the long gaps that occur each
year when the Sun passes too close to the source.  The autocorrelation
function was calculated on the interpolated grid.  Since the source
sampling is usually at approximately 3.5~dy, these plots are only
informative at times longer than that.

In Table~2 we present three different measures of the percentage of
time that each source is active.  Considering the range of behaviours
and luminosities among the sources, no single measurement of what it
means for a system to be ``active'' or ``in outburst'' is universally
sufficient, nor do we expect one of these three definitions to be
useful for every purpose.  The first measure is the percentage of time
spent with a flux of at least three standard deviations above zero;
the standard deviation in this case is defined by averaging the lowest
20 per cent of the error bars from all the bulge scan fits.  These
errors are partly statistical and partly based on fitting
uncertainties from the influence of other nearby sources.  In general,
the lowest 20 per cent of the errors correspond to times when the
source is undetectable to the PCA, with the exception of the sources
that are almost always visible, such as 4U~1700--377.  The second
measure is the percentage time that the source is above
40~counts~s$^{-1}$ (2--10~keV, five detectors equivalent), giving a
constant threshold in observed flux.  The last measure is the
percentage of time spent at a luminosity above 1$\times
10^{35}$~erg~s$^{-1}$ at the first listed distance to each
source shown in Table~1.  Bulge scan count rates (2--10~keV) are
converted to energy flux (also 2--10~keV) by assuming a thermal
bremsstrahlung spectrum with temperature 20~keV, typical of SFXTs, and
using the WebPIMMS tool at NASA's High Energy Astrophysics Science
Archive Research Centre (HEASARC)
(http://heasarc.nasa.gov/Tools/w3pimms.html, retrieved 18 April 2011).
Any of these can be used as a simplified activity parameter, but more
information is contained in the full luminosity histograms (panel~c of
Fig.~1--8).

Table~2 also gives the average and maximum luminosity of each source,
the average value for the bottom half of luminosity measurements, and
the average of the bottom 20 per cent of luminosity uncertainties.
The latter number serves two purposes: first, to make it clear when
the average of the bottom half of luminosity is significantly above
zero (only for 4U~1700--377 and XTE~J1901+014), and second, to give
the luminosity threshold used for the first activity index: e.g., 
XTE~J1739--302 spent 4.8 per cent of its time with a flux above 
$3\times(2.30\times10^{34}$~erg~s$^{-1})$.
The negative average luminosities for the bottom half of
observations in IGR J17544--2619 and IGR J16479--4514 are due to the
statistical and systematic uncertainties in the background-subtracted
PCA count rates.  Negative values in these cases just indicate that
in a typical measurement, the uncertainties due to subtracting both
instrumental background and the influence of other sources are
greater than the typical count rate of the source.

\citet{romano11}
showed luminosity histograms for three sources (differential instead
of integral), but summarised these results with an {\it inactivity}
measurement instead.  That was most appropriate for their data set,
which is only 2 years long -- and therefore contains fewer outbursts
-- but very sensitive, since the {\it Swift} XRT is a true imaging
instrument.  Thus they have good discrimination between low and very
low emission levels.  The {\it RXTE} PCA bulge scan data are not as
easy to interpret at the lowest fluxes, even though PCA is also very
sensitive, due to complications in interpreting the scan data due to
the structure of diffuse X-ray emission in the Galactic ridge.  On the
other hand, the PCA data set contains over 4000 pointings covering
over 12~yr for some of the sources we include, so that our database
is the most sensitive to the distribution of the rarer high fluxes.

\begin{table*}
 \centering
  \caption{System characteristics for eight fast transients}
  \begin{tabular}{@{}llll@{}}
  \hline
   Name & Companion & Distance & Period \\
 \hline
XTE~J1739--302 & O8Iab(f) (1,2) & 2.7 (2) & 51.47 $\pm$ 0.02 (3)\\
IGR~J17544--2619 & O9Ib (4) & 3.6 (2,4) & 4.926 $\pm$ 0.001 (5) \\
4U~1700--377 & O6.5Iaf+ (6) & 1.9, 2.1 (6,7) & 3.41161  $\pm$ 0.00005 (8) \\
SAX~J1818.6--1703 & $\sim$B0I (9,10) & 2, 2.1 (10,11) & 30.0 $\pm$ 0.2 (12,13)\\
AX~J1845.0--0433 & O9Ia (14) & 7 (14) & unknown \\
IGR~J16479--4514 & O8.5I, O9.5Iab (2,15) & 4.9, 2.8 (2,15) & 3.3194 $\pm$ 0.001 (16,17)\\
XTE~J1743--363 & unknown & unknown & unknown \\
XTE~J1901+014 & faint in {\it K} (10) & 1--7 kpc (10,18) & unknown \\
\hline
\end{tabular}

(1) Negueruela et al. 2006;
(2) Rahoui et al. 2008;
(3) Drave et al. 2010;
(4) Pellizza, Chaty, \& Negueruela 2006;
(5) Clark et al. 2009;
(6) Ankay et al. 2001;
(7) Megier et al. 2009;
(8) Hong \& Hailey 2004;
(9) Negueruela \& Schurch 2007;
(10) Torrej\'{o}n et al. 2010;
(11) Negueruela, Torrej\'{o}n, \& Reig 2008;
(12) Zurita Heras \& Chaty 2009;
(13) Bird et al. 2009;
(14) Negueruela et al. 2007;
(15) Nespoli, Fabregat, \& Mennickent 2008;
(16) Jain, Paul, \& Dutta 2009;
(17) Romano al. 2009;
(18) Karasev, Lutovinov \& Burenin 2008
\end{table*}

\begin{table*}
 \centering
  \caption{X-ray luminosity parameters for eight fast transients}
  \begin{tabular}{@{}lrrrrrrr@{}}
  \hline

Name & Per cent & Per cent  & Per cent & Average$^a$ & Average, lower & rms, lower & Maximum$^a$ \\ 
 &$> 3\sigma$ & $> 40$ c/s &$> 1 \times 10^{35}$~erg~s$^{-1}$ & & 50 per cent$^a$ &20 per cent$^a$ &  \\ 
\hline\\
XTE J1739--302 &   4.8 &   4.1 &   2.3 &   2.25 &  -0.15 &   2.30 &  115.86\\
IGR J17544--2619 &   5.8 &   2.6 &   3.6 &   1.25 &  -2.10 &   2.41 &  156.58\\
4U 1700--377 &  85.3 &  84.4 &  77.1 &  86.79 &  15.91 &   1.16 & 1240.55\\
SAX J1818.6--1703 &  26.4 &  15.2 &   3.1 &   2.58 &   0.75 &   0.96 &   39.40\\
AX J1845.0--0433 &  35.5 &   6.5 &  68.4 &  27.00 &   5.56 &   7.10 &  384.19\\
IGR J16479--4514 &  11.2 &  42.3 &  56.6 &  22.80 &  -9.21 &  23.59 &  193.66\\
XTE J1743--363 &  50.6 &  10.0 &  31.4 &  10.36 &   1.97 &   1.97 &  297.95\\
XTE J1901+014 &  83.9 &   1.9 &   0.6 &   5.20 &   3.85 &   1.13 &   10.90\\
\hline\\
\end{tabular}

$^a$ {\it luminosity in units of $10^{34}$ erg s$^{-1}$  at the
first listed source distance in Table~1, or 3 kpc for XTE J1743--363
and XTE J1901+014}.

\end{table*}

\section{Results and Discussion}

\subsection{XTE~J1739--302}

\citet{drave10} recently reported evidence for a 51.47~d period in
this system, based on two separate analyses of {\it INTEGRAL}/IBIS data.  In
the first, they performed a Lomb-Scargle periodogram on the full IBIS
light curve and found the highest power at this period, with a
significance greatly exceeding 99.999 per cent confidence.  They also
collected a number of outbursts from IBIS, {\it Swift}, and {\it ASCA}
data and plotted their frequency of occurrence and intensity with
respect to orbital phase, finding that most outbursts happened around
phase 0.5 according to their ephemeris with zero phase defined 
arbitrarily at MJD 52698.2.

We find no evidence for any power at this period in our periodogram
(Fig.~1b), but \citet{drave10} point out that this was also true of
{\it Swift} and {\it RXTE} ASM data.  We note that the PCA bulge scan
data presented here have a longer baseline than either {\it Swift} or
{\it INTEGRAL} and a much higher sensitivity than the {\it RXTE} ASM.
The twice-weekly monitoring means that the 51.47~d period is very well
sampled throughout the 11 years.  

\begin{table}
 \centering
  \caption{Outbursts of XTE~J1739--302 not included by \citet{drave10}}
  \begin{tabular}{@{}llll}
  \hline
   MJD & Phase & Major Outburst? & Source \\
 \hline

 50672.7 & 0.647 &Y & {\it RXTE} pointed \\
 51248.3 & 0.830 &Y & {\it ASCA} \\
 51849.2 & 0.505 &Y & {\it RXTE} pointed  \\
 51977.9 & 0.005 &Y & {\it RXTE} scan \\
 52452.9 & 0.234 &Y & {\it RXTE} pointed  \\
 52888.3 & 0.693 &Y & {\it INTEGRAL}/IBIS \\
 53569.7 & 0.932 &Y & {\it RXTE} scan \\
 51265.8 & 0.170 & & "  \\
 51338.5 & 0.582 & & "  \\
 51660.8 & 0.844 & & "  \\
 51677.7 & 0.173 & & "  \\
 52107.5 & 0.524 & & "  \\
 52560.3 & 0.321 & & "  \\
 52779.7 & 0.583 & & "  \\
 52945.1 & 0.796 & & "  \\
 53140.5 & 0.594 & & "  \\
 53301.8 & 0.727 & &"  \\
 53646.1 & 0.416 & & " \\
 53649.1 & 0.475 &Y & " \\
 53781.0 & 0.038 & &"  \\
 53784.9 & 0.114 & &"  \\
 53801.8 & 0.442 & &"  \\
 53847.8 & 0.335 & &"  \\
 53858.6 & 0.545 & & " \\
 53861.5 & 0.602 & & " \\
 53995.2 & 0.200 & &"  \\
 54584.8 & 0.655 & &"  \\
 54686.9 & 0.637 & &"  \\
 54719.2 & 0.266 & &"  \\
 54929.8 & 0.357 & & " \\
 54939.8 & 0.551 & & " \\
 54954.6 & 0.840 &Y &"  \\
 55052.6 & 0.742 & &"  \\
 55085.7 & 0.387 & &"  \\
 55088.6 & 0.443 & &"  \\
 55286.3 & 0.284 & &"  \\
 55395.6 & 0.408 & &"  \\
 55439.9 & 0.268 & &"  \\
 55453.6 & 0.533 & &"  \\
 55460.2 & 0.663 & &"  \\
 53581.0 & 0.152 & & {\it Swift} \\
 53663.5 & 0.755 & & " \\
 53798.2 & 0.372 & & " \\
 54140.7 & 0.026 & &"  \\
 54632.3 & 0.577 & & " \\
 54673.6 & 0.380 & &"  \\
 54724.5 & 0.369 & &"  \\
 54900.7 & 0.792 &Y & " \\
\hline
\end{tabular}

\end{table}

Our result does not, however, necessarily constitute a contradiction
of the \citet{drave10} result.  Each PCA measurement is a snapshot
lasting only a few seconds as the spacecraft scans across the source.
Most {\it INTEGRAL} measurements used by \citet{drave10}, on the other
hand, are full 2000~s science windows of that instrument.  If we
hypothesise that the typical behaviour of the source near periastron
is to produce small flares that tend to come at least once every
2000~s but with a duty cycle of $<< 50$ per cent, then it would be plausible
that the PCA should miss making positive detections at many periastron
transits while {\it INTEGRAL} might catch most of them.

We therefore turn to the other technique used by \citet{drave10}, the
identification of the phase of medium-to-large outbursts according to
the ephemeris they deduced from their periodogram.  We have identified
48 outbursts that were not included in the sample given by
\citet{drave10}.  These include events from the bulge scans presented
here, including some that were presented in \citet{smith06}, outbursts
observed during PCA pointings to 1E~1740.7--2942 given in the same
work, {\it Swift} outbursts from \citet{romano09}, and one from {\it
INTEGRAL}/IBIS that was not included by \citet{drave10}.  We also
include the major outburst observed by {\it ASCA} in 2002
(MJD~51248.3) \citep{sakano02}, which was incorrectly assigned a phase
of 0.468 by \citet{drave10}, near the apparent periastron,
but was not used in the histogram in their fig.~4, which included
35 events from {\it INTEGRAL}/IBIS alone.  The correct phase of the
{\it ASCA} outburst is 0.83.  In many cases, more than one spacecraft
observed what was probably the same outburst.  Since there is no
prescribed way to distinguish one long outburst from two separate
ones, we somewhat arbitrarily discarded all new outbursts that were
within 1~dy of one of those used by \citet{drave10}, and similarly
made sure that no two of the new data points were within 1~dy of each
other.  The new events are given in Table~3.

\begin{figure}
\includegraphics[width=8cm]{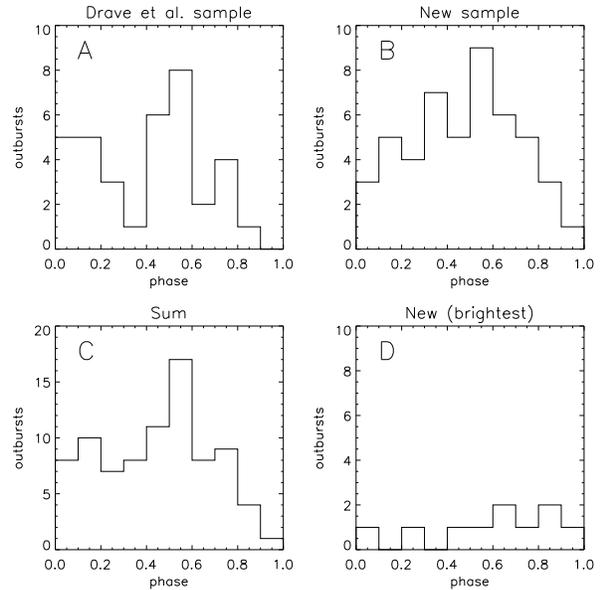}
\caption{Histograms of outburst phases for XTE~J1739--302 according
to the ephemeris of \citet{drave10}. A: outbursts reported
by \citet{drave10}.  B: new sample (see text and Table~3).
C: the sum of panels~A and B.  D: the brightest
ten events in the new sample. }
\end{figure}

Fig.~9 shows histograms of outburst phase for the original sample of
events used by \citet{drave10}, the new events, and the sum.  The
histogram in Fig.~9a is identical to that given in fig.~4 of
\citet{drave10}.  The new sample, in Fig.~9b, also peaks near 0.5 in
phase.  To compare these results to the null hypothesis, we calculated
the sum of the squares of the deviations from the average for each
plot and compared it to the same quantity for 10$^6$ simulations of
the null hypothesis with the same number of events.  The real data
sets from panels~A, B, and C had a sum 
greater than or equal to all but 5.6, 40.4, and 2.4 
per cent, respectively, of the corresponding simulations.  This shows
that while the new data set provides no evidence for periodicity in
its own right, it does increase, and not decrease, the significance of
the original result when the samples are combined.  Fig.~9d
shows the phase histogram of the ten brightest outbursts in the new
sample, corresponding to fluxes over 6$\times
10^{-10}$erg~s$^{-1}$cm$^{-2}$ or 5$\times 10^{35}$erg~s$^{-1}$ at
2.7~kpc and marked ``Major'' in Table~3.  There is no apparent
clustering in phase for these events, which include the {\it ASCA}
outburst at phase 0.83, near the minimum occurrence phase for most
outbursts.

The integral luminosity distribution at high fluxes (Fig.~1c) has a
more or less power-law-like distribution (index of approximately
--1.5), a characteristic shared with
IGR~J17544--2619 (Fig.~2c), the other ``canonical'' SFXT system, and
SAX~J1818.6--1703.  The autocorrelation function (Fig.~1d) shows 
no significant signal.  If outbursts occur randomly and last for less
than the average sampling period (3.5~dy), this is to be expected.

\subsection{IGR~J17544--2619}

The orbital period reported by \citet{clark09} at $4.926 \pm 0.001$~d
appears clearly in the PCA data (Fig.~2b) at the same period and with
comparable error.  This signature also appears to a lesser extent in
the autocorrelation data (Fig.~2d), where it might be taken for noise
if the period were not already known.  The intensity distribution and
lack of significant autocorrelation of nearby observations look very
similar to XTE~J1739--302.  If there is indeed an order of magnitude
difference between the orbital periods of XTE~J1739--302 and
IGR~J17544--2619, it is rather extraordinary that their behaviours are
so similar, in frequency of outbursts (shown here), frequency of
inactivity \citep{romano11}, and in
the detailed appearance of the outbursts themselves
\citep[e.g.]{sidoli09}. The optical counterparts of IGR~J17544--2619
\citep*{pellizza06} and XTE~J1739--302 
\citep{negueruela06,rahoui08} are very similar as well, which argues
against an expectation that
large differences in the structure and intensity of their
winds can compensate for the different orbital radii, producing
similar X-ray behaviour.  Models that rely on unusual magnetic 
properties of the neutron star to explain SFXT outbursts
\citep*{grebenev07, bozzo08a, li11} should be studied further to see if they can
produce uniform variability behaviour under greatly differing wind
conditions.

\subsection{4U~1700--377}

This persistently bright wind accretor is shown as a point of comparison
for the more transient systems. Its maximum luminosity is higher than
that of the two canonical SFXTs discussed above by factors of 11 and 8,
but the {\it average} luminosity is higher by factors of 39 and 69.
The comparison of the average luminosity is probably very good, but 
comparing the maximum luminosity is harder, since we don't know for
how long each system tends to stay at this maximum in each outburst, 
or how common they are.  The true maximum level within this 11-year
period may have been missed in one system and caught in another.  For
example, we know from coverage of one outburst of IGR~J17544--2619
by {\it Suzaku} \citep{rampy09} that the system reached at least $5.3\times
10^{36}$erg~s$^{-1}$.  This is as bright as all but one of the data
points from the bulge scans of 4U~1700--377.  The fast transients
are capable of reaching almost the maximum luminosity of this
bright system, but spend most of their time at low luminosities. 

The well-known orbital period of 4U~1700--377 (see Table~1) is clearly
visible, and the autocorrelation function shows coherence on a time-scale 
of $\sim30$~dy.  This could be due to a process intrinsic to the
companion star producing slow variations in its wind, a feedback
between the accretion luminosity of the neutron star and mass loss
from the companion, or the presence of an accretion disc around the
neutron star, with a viscous time-scale of this order.

\subsection{SAX~J1818.6--1703}

This system \citep{zand98} looks similar to the two canonical SFXTs in terms of
the power-law shape of the distribution of luminosities (Fig.~4c).
The orbital period (See Table~1) and its harmonics are clearly 
seen in the periodogram; even the autocorrelation function shows
a correlation at the orbital period, which we believe represents
the high probability of an unrelated outburst at the next periastron
passage.  \citet{sidoli09} found that the outbursts of this system,
studied in depth with {\it Swift}, show very similar characteristics
to the canonical SFXTs.  Its 30~dy orbital period is approximately
halfway between those of IGR~J17544--2619 and XTE~J1739--302, showing again
that very similar outburst duty cycles and luminosities are seen with
different orbital parameters.
While the system has been shown to have as
deep a quiescence as the canonical SFXTs \citep{bozzo08b}, there has
not been enough monitoring by a sensitive telescope such as {\it Swift}
to get the good data on frequency of quiescence that are now available
for other systems \citep{romano11}.

\subsection{AX~J1845.0--0433 and IGR~J16479--4514}

There was only one major and one minor outburst of AX~J1845.0--0433
caught by the 
bulge scans, as can be seen in Fig.~5, panel~(a), and no obvious outburst
of IGR~J16479--4514.  The contamination of AX~J1845.0--0433 by
Galactic plane diffuse emission, visible as a bifurcation between those
scans taken perpendicular and parallel to the plane, makes any conclusion
about the flux outside of outbursts difficult; the contamination of
IGR~J16479--4514 by GX~340+0 produces a similar problem.  While both
of these sources show a fairly high average luminosity in the fifth
column of Table~2, we do not claim this result as significant; 
repeated observations with a focusing instrument such as the {\it Swift}
XRT would be preferable to establish the quiescent and low-flux behaviour
of these sources.

\subsection{XTE~J1743--363}

\subsubsection{Position and identification}

\begin{figure}
\includegraphics[width=8cm]{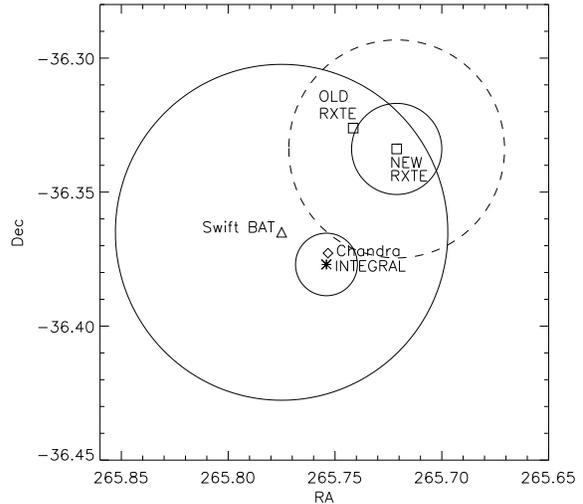}
\caption{Summary of position measurements for XTE~J1743--363.  The
{\it Chandra} position has too small an error circle to be plotted on
this scale, and the first {\it RXTE} position has too large a circle.
The new {\it RXTE} position is shown with the 3$\sigma$ contour (dashed
line) as well as the 1$\sigma$ contour (solid line).
See text.}
\end{figure}

This source was discovered by \citet*{markwardt99}.  Its occasional
fast outbursts have led to the suggestion that it might be a SFXT, and
the X-ray spectrum in outburst could be fit with a thermal
bremsstrahlung spectrum with temperature 22~keV, typical of X-ray
pulsars and SFXTs \citep{sguera06}.  \citet{ratti10}, using a {\it
Chandra} observation of the field (2008~Feb~8, principal
investigator M. Mendez), identified a cluster of 11 X-ray photons
corresponding to the position of the bright infrared source
2MASS~17430133--3622221.  This is the only significant source in the
{\it Chandra} field, although it is possible that XTE~J1743--363 was
entirely quiescent at that time and the X-rays come from another
system.  Fig.~10 compiles some X-ray position measurements for
XTE~J1743--363: the initial {\it RXTE} estimate at the source's
discovery \citep{markwardt99} (at 10', its error circle is too large
to appear on the plot), the most precise {\it INTEGRAL}/IBIS position
(0.7' error circle, 90 per cent confidence) \citep{bird10}, the {\it Swift}
BAT position from the 58-month catalogue \citep{baumgartner10}, the {\it
Chandra} position \citep{ratti10}, and our new, refined {\it RXTE}
position.  Earlier {\it Swift}/BAT positions, while different, are
consistent with both the newest BAT position and the {\it Chandra}
source \citep{tueller10,cusumano10}.

The {\it RXTE} position was derived from a dedicated scan of the
source using the PCA instrument.  The observation was on 1999~Feb~15
at 01:15~UT for 2.4~ksec.  The scan pattern consisted of a connected
path, including scans along constant right ascension and declination.
The model included XTE~J1743--363 and one other known source in the
field of view, 4U~1746--37, at a fixed position.  The primary change
in deriving the refined position was that the intensity of
XTE~J1743--363 was allowed to vary with a long-term linear trend.  The
uncertainties of individual measurements were increased in order to
de-weight the strong variability of the source, which was near 35 per cent
rms.  While the {\it Chandra} position is slightly outside the
3$\sigma$ {\it RXTE} contour, the susceptibility of the {\it RXTE}
fit to variability, combined with the agreement of {\it INTEGRAL}/IBIS with
{\it Chandra} (spanning both the low- and high- energy ends of
the PCA energy range), leads us to conclude that all the observatories
are seeing the same source.

\subsubsection{X-ray characteristics}

\begin{table*}
 \centering
  \caption{List of PCA pointings to XTE~J1743--363 and absorbed bremsstrahlung fits}
  \begin{tabular}{@{}llrrrr@{}}
  \hline
Observation ID & MJD & Rate$^a$ & nH$^b$ & kT (keV) & Normalisation$^c$ \\
  \hline
40408-01-01-00 & 51224.051 & 12 & 11.05 $\pm$ 0.22 & 13.76 $\pm$ 0.35 & 5.509 $\pm$ 0.082\\
40408-01-02-00 & 51230.781 & 22 &  6.11 $\pm$  0.16 & 13.67 $\pm$ 0.34 & 8.24 $\pm$ 0.11\\
50138-02-01-00 & 51706.236 & 23 & 17.40 $\pm$  0.43 & 14.66 $\pm$ 0.54 & 11.44 $\pm$ 0.27 \\
50138-02-02-00 & 51707.631 & 13 & 10.55 $\pm$  0.51 & 15.74 $\pm$ 0.97 & 4.64 $\pm$ 0.15\\
50138-02-03-00 & 51710.759 & 11 & 10.26 $\pm$  0.66 & 11.42 $\pm$ 0.77 & 4.33 $\pm$ 0.20\\
50138-02-04-00 & 51713.016 & 11 & 8.77  $\pm$  0.53  & 10.54 $\pm$ 0.58 & 4.01 $\pm$ 0.16 \\
92047-04-01-02 & 54115.404 & 2.0 & -- & -- & -- \\
92047-04-01-04 & 54115.461 & 2.0 & -- & -- & -- \\
92047-04-01-00 & 54115.520 & 1.9 & -- & -- & -- \\
92047-04-01-01 & 54115.910 & 1.9 & -- & -- & -- \\
92047-04-02-00 & 54301.307 & 5.3 & 11.6 $\pm$   1.4  &  14.5 $\pm$ 2.2   &  1.57 $\pm$  0.14\\
92047-04-02-01 & 54303.272 & 3.2 & 7.4 $\pm$   1.7  &  15.9 $\pm$ 3.8   &  0.503 $\pm$  0.076\\
92047-04-02-02 & 54304.383 & 3.2 & 8.4 $\pm$   1.6  &  16.4 $\pm$ 3.7   &  0.550 $\pm$  0.075\\
 \hline\\
\end{tabular}

$^a$ {\it counts/s/PCU}.

$^b$ $\times 10^{22}$ cm$^{-2}$. 

$^c$ $\times 10^{-2}$ as defined in equation (1).
\end{table*}

\begin{table*}
 \centering
  \caption{Absorbed bremsstrahlung plus blackbody fits to XTE~J1743--363 spectra}
  \begin{tabular}{@{}llrrrr@{}}
  \hline
Observation ID &  nH$^a$ & Bremss. kT (keV) & Bremss. norm.$^b$ & Blackbody kT (keV) & Blackbody norm.$^c$ \\
  \hline
40408-01-01-00 &   8.52 $\pm$  0.47 &   29.5 $\pm$  6.7 &  3.09 $\pm$ 0.30 &  1.673 $\pm$ 0.042 &  0.79 $\pm$ 0.23 \\
40408-01-02-00 &   3.69 $\pm$  0.42 &  23.5 $\pm$  4.5  &  4.67 $\pm$ 0.50 & 1.801 $\pm$ 0.050 &  1.21 $\pm$ 0.33 \\
50138-02-01-00 &   13.44 $\pm$  0.86 &  24.9  $\pm$ 6.2 &   6.15 $\pm$ 0.89 & 2.058 $\pm$ 0.085&  1.82 $\pm$ 0.54  \\
 \hline\\
\end{tabular}

$^a$ $\times 10^{22}$ cm$^{-2}$.

$^b$ $\times 10^{-2}$ as defined in equation (1).

$^c$ $\times 10^{-3}$ normalised to 10$^{39}$ erg s$^{-1}$ at 10~kpc \citep{arnaud11}.
\end{table*}

\begin{figure}
\includegraphics[width=0.45\textwidth]{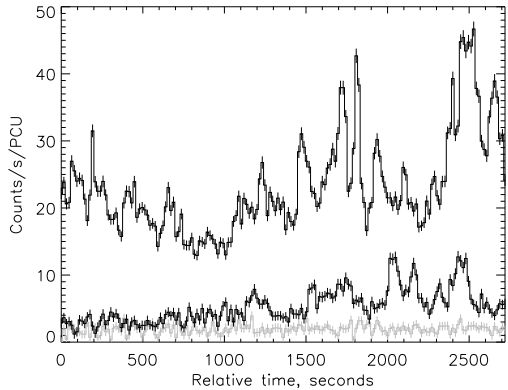}
\caption{X-ray variability at three representative luminosities of XTE~J1743--363.
Top trace: observation 50138-02-01-00.  Middle trace: observation 92047-04-02-00.
Bottom trace (grey): observation 92047-04-01-00.  See Table~4.}
\end{figure}

\begin{figure}
\includegraphics[width=0.45\textwidth]{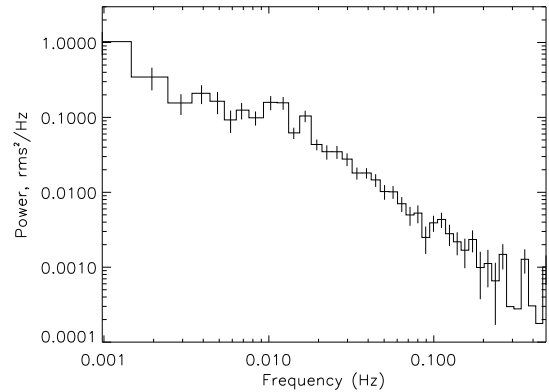}
\caption{
Power spectrum combining the data from observations 40408-01-02-00, 50138-02-01-00,
50138-02-02-00, 50138-02-03-00, and 50138-02-04-00.  No pulse period is visible between
2~s and 1024~s.}
\end{figure}

In hopes of getting further insight into the nature of this source, we
examined twelve archived {\it RXTE}/PCA pointings, some from 2000, when it
was often bright, and some from 2007, when it was much less active
(see Fig.~7).  Table~4 shows the dates, {\it RXTE} observation
identification numbers, and count rates for these pointings.
Fig.~11 shows some sample light curves and Fig.~12 shows the power
spectrum made by averaging the power spectra from five of the observations.

Four observations made on MJD 54115 showed low and nearly identical
count rates of 1.9--2.0 counts s$^{-1}$ and no variability. The flux
varied during all the other observations on time-scales of seconds to
minutes.  We used {\sc xspec} version 12.5.1 to fit the
PCA spectra.  These four quiet observations included an iron line of
high equivalent width, and no detectable interstellar absorption above
3~keV, in contrast to all the other observations, which showed
significant and variable absorption.  Finally, the count rates of the
four spectra from MJD 54115 are comparable to the flux level that is
identified as diffuse from the PCA bulge scans in this region.  Taking
all this evidence together, we believe it quite likely that during
these four observations we were seeing only the Galactic diffuse
component and not the source itself.  The best-fitting continuum
spectrum to these four observations together was an optically thin
thermal bremsstrahlung spectrum of temperature $10.10 \pm 0.50$~keV
and normalisation $(5.99 \pm 0.16)\times 10^{-3}$, defined as

\begin{equation}
\frac{3.02 \times 10^{-15}}{4\pi D^2} \int n_e n_I dV
\end{equation}
where D is the distance to the source in centimetres and $n_e$ and
$n_I$ are the electron and ion densities per cubic centimetre at the
source \citep{arnaud11}.  An iron line of equivalent width 0.43~keV
and centre energy ($6.63 \pm 0.54$)~keV was also required (at
3.7$\sigma$), and was consistent with being narrow, although the width
was poorly constrained ($<$2.9~keV).  In fitting the rest of the spectra,
we included these two spectral components with no free parameters.
No further iron line was needed in any of the other fits.

In Table~4 we show the spectral parameters of the remaining flux,
presumed to be from XTE~J1743--363 itself.  For most of the spectra,
an absorbed optically thin thermal bremsstrahlung spectrum provided a
good fit.  We used the Tuebingen-Boulder ISM absorption model
({\sc tbabs}) \citep*{wilms00}.  For the three spectra with the
best counting statistics, however, an additional component was
necessary to obtain $\chi$-square values consistent with chance.  A
thermal blackbody spectrum served this purpose well for all three
spectra, as it typically does for accreting neutron stars.  Table~5
shows the resulting fit parameters in this case.  Including the
blackbody component has the effect of raising the temperature of the
bremsstrahlung component and reducing the best-fitting absorption
column.

It is apparent from Tables~4 and 5 that the absorption column is both
high and variable, and therefore local to the system.  This suggests
a companion with a dense wind, which can be characteristic of both
blue supergiants and red giants such as AGB stars.  The absorption
column doesn't seem to correlate with luminosity at all, suggesting
that the absorption is taking place in foreground structures in the
wind rather than in the concentration of material accreting on to the
compact object at the moment.

\subsubsection{Optical spectroscopy}

\begin{figure}
\includegraphics[width=9cm]{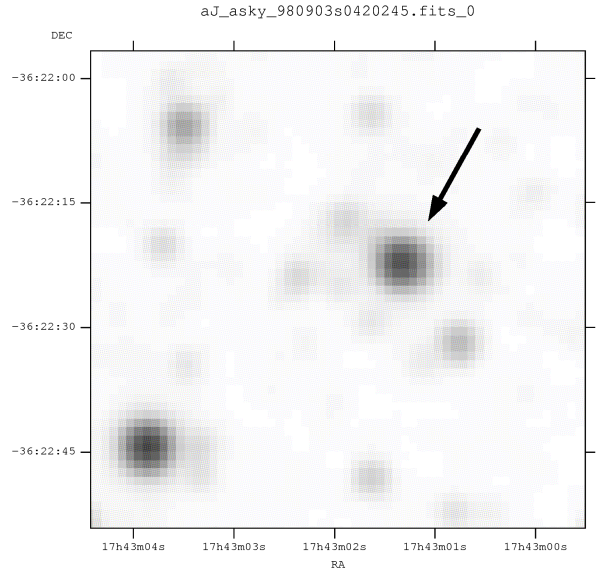}
\caption{2MASS {\it J}-band image of the region around 2MASS~17430133--3622221.
The arrow marks the source.}
\end{figure}

\begin{figure}
\includegraphics[width=8cm]{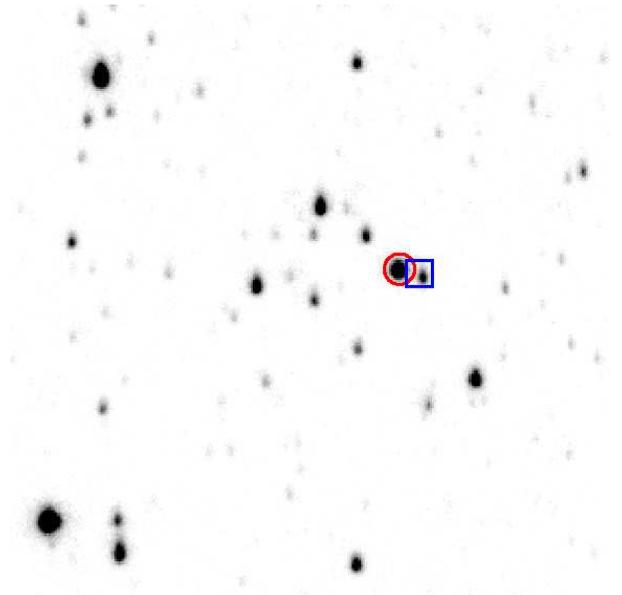}
\caption{NOT finder image covering the
same area as Fig.~13.  Note that 
2MASS~17430133--3622221 is resolved into two sources.  The brighter
source to the left (marked with a circle) is the M giant discussed
in the text.}
\end{figure}

\begin{figure*}
\includegraphics[width=0.5\textwidth,angle=270]{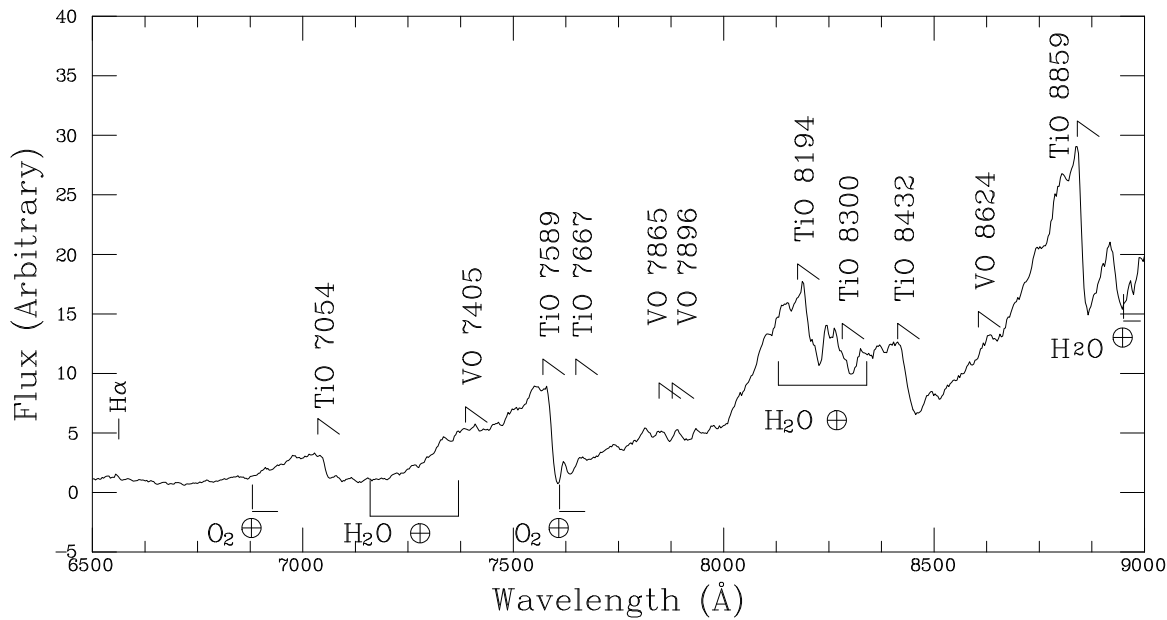}
\caption{ALFOSC spectrum of 2MASS~17430133--3622221/USNO--B1.0~0536--0518280.
The positions of some relevant molecular bandheads are indicated.  The
strongest telluric features are also marked. }
\end{figure*}

\citet{ratti10} used the infrared colours of 2MASS~17430133--3622221
to constrain the spectral type of the companion, but not severely;
they found that a giant of type G0--6, a G or K supergiant, or a
supergiant earlier than O9 were all consistent with the 2MASS colours,
assuming varying degrees of reddening.  The 2MASS magnitudes are {\it J}=9.616,
{\it H}=8.305, and {\it K}=7.624.  The 2MASS source (Fig.~13) is coincident with
USNO--B1.0~0536--0518280, with  {\it R=16.4} and 16.77 (R1 and R2
measurements) and {\it I}=13.78 (not seen in the USNO {\it B} band) and with the
Midcourse Space Experiment infrared source MSX6C G353.3697-03.4195,
seen only in band {\it A} at 0.1386~Jy.  Band {\it A} is centred at
8.28~$\mu$m with half-maximum points at 6.8 and 10.8~$\mu$m 
according to \citet{egan03}.

We obtained a far-red spectrum of the counterpart using the
Andalucia Faint Object Spectrograph and Camera (ALFOSC) on
the 2.6~m Nordic Optical Telescope (NOT) in La Palma, Spain, on the
night of 2009 June 27. The instrument was equipped with
a thinned $2048\times2048$ pixel E2V CCD, covering a field of
view of $6\farcm4\times6\farcm4$ with a spatial scale
$0\farcs19$/pixel. We used grism \#4 with a $1\farcs0$ slit, providing a
resolving power $R\approx450$. In order to remove fringing, an internal
flat was taken at the target position.

Because of the Southern declination of the target, it was observed at
a very high airmass ($Z=2.4$). However, as the finder image (Fig.~14)
shows, image quality was good. 2MASS~J17430133$-$3622221 can be
clearly resolved into two objects, separated by $\sim2\arcsec$, both
of which fell within the slit.  The fainter object is a normal G/K
star. The brighter star is a very red object, and the obvious
counterpart to the 2MASS source. Its spectrum (Fig.~15) is dominated
by deep molecular bands. In particular, we see deep bandheads of TiO,
typical of M-type stars.

\citet{negueruela11} have shown that the depth of the TiO 8859\AA\
bandhead is a very good indicator of effective temperature. Though the
resolution of our spectrum is too low to use it as a $T_{{\rm eff}}$
calibrator, the fact that its depth is $\ga0.5$ of the continuum
immediately shortward indicates that it is not a dwarf star. Only giants
and supergiants present such deep molecular features. It also indicates
that the spectral type is later than M6. This is confirmed by the
detection of VO bandheads, which are only visible at spectral types M7 and
later \citep{turnshek85,gray09}.

The deep depression of the continuum in the 7400\,--\,7550\AA\ and
7860\,--\,8050\AA\ regions can only be due to the presence of strong
VO bands of the extreme red system, as there are no telluric features
in these two regions. Moreover, the VO bandhead at 8624\AA\ is also
likely detected. Though a better spectrum is needed to obtain an
accurate spectral type, the data are not very different from the
expectation for spectral type M8.
As supergiants with such a late spectral type are very rare
\citep{levesque2005}, the object is very likely a giant. 

The fundamental properties of such late stars are not well calibrated.
For M8\,III stars, \citet{vanb99} estimate $T_{{\rm eff}}=3050$~K,
though this is an extrapolation from earlier spectral types. Their fit
to the intrinsic $(V-K)_{0}$ colour indicates values between
$(V-K)_{0}=-8.0$ and $(V-K)_{0}=-8.5$ at this temperature. As
\citet{the90} find $M_{V}=+1.6$ on average for M8\,III giants, their
infrared absolute magnitude is $M_{K}\la-6.4$.  Late M-type giants may
be close to tip of the red giant branch (RGB) or on the asymptotic
giant branch (AGB).  Many giants later than M6 are brighter than the
tip of the RGB \citep[e.g.,][]{tabur09a}, which has observationally
been set at $M_{K}=-6.85$ \citep{tabur09b}, and so most M8\,III
giants are likely on the AGB.

\begin{figure*}
\includegraphics[width=0.9\textwidth]{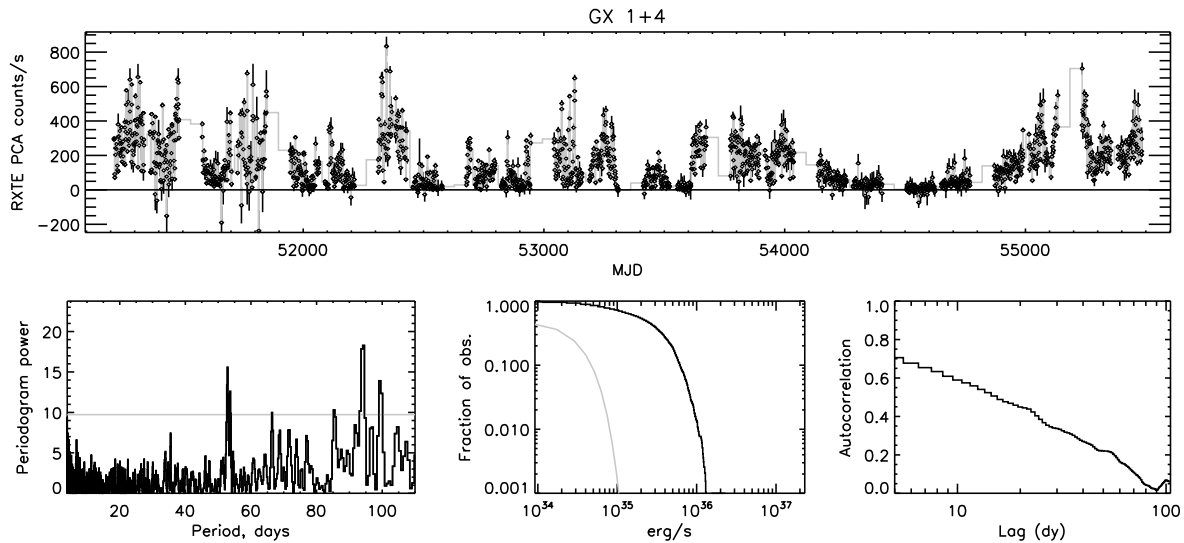}
\caption{Data products as described in Fig.~1, for
the symbiotic binary GX~1+4.}
\end{figure*}
 
Theoretical magnitudes calculated from {\tt PHOENIX} models and calibrated
against observational data suggest that all late-M giants have
$(J-K)_{0}\sim1.3$ \citep{kuc05}, in good agreement with
extrapolation of the colour calibration of \citet{sl09}. The calibration of
\citet{groene04}, based on older models, gives $(J-K)_{0}=1.37$ for
dust-free M8 stars. This reference also gives $(J-K)_{0}=1.97$ for
dust-enshrouded M7 stars. Based on this, our target cannot be
dust-enshrouded, unless the insterstellar extinction is negligible. This
possibility is ruled out by the non-detection in all MSX bands except for
A, and so we assume that the star is not surrounded by dust. Therefore its
colour excess is $E(J-K)_{0}\approx0.6-0.7$. For a standard reddening law,
$A_{K}\approx0.4-0.5$ and so the distance modulus should be
$DM\approx13.5$, i.e., $d\approx5$~kpc.  As observations of LMC giants
suggest that $M_{K}$ may deviate by $\pm0.6$~mag from the average at a
given $(J-K)$ \citep{cioni01}, we may settle for  $d\approx5\pm1$~kpc,
though there are many other sources of uncertainty not considered in this
error estimate.

For instance, these calculations assume that the star has characteristics
typical of M-type giants in the solar neighborhood. Our upper distance
limit, however, brings us close to the distance to the Galactic bulge. At
$b=-3\fdg4$, the extinction to our source is not incompatible with
Galactic bulge membership \citep{dutra03}, even if we take into account
the possibility of bluer colours for bulge M-giants \citep{frogel}.
However, for a  distance of 8.5~kpc, the source would have $M_{K_{{\rm
S}}}\approx-7.8$, only compatible with the very brightest bulge M-type
stars \citep{gs02}. Our target would then have to be a thermally-pulsating
AGB.

Even in the most likely case that the star is foreground to the bulge, it
is likely to be variable, as most late-type giants are long-period
variables. Their amplitude in the {\it K} band is generally small and
unlikely to dominate our error budget. On the other hand, photometric
determination of a pulsation period would be the most direct way to
evaluate the stellar luminosity and hence distance.

\subsubsection{XTE~J1743--363 as a symbiotic X-ray binary}

The spectral identification of the companion to XTE~J1743--363 as a
late M giant puts the system in the small group of symbiotic X-ray
binaries, although we have not seen an additional hot component to the
optical spectrum, which was the origin of the term ``symbiotic''.  For
comparison, we show in Fig.~16 the PCA bulge scan data for GX~1+4, a known
symbiotic X-ray binary in which an X-ray pulsar orbits an M giant.
\citet{hinkle06} used infrared spectroscopy to derive an orbital
period of 1161~dy, a luminosity of 1270~$L_{\sun}$, and a distance of
4.3~kpc, which we assume to scale the luminosity distribution in
Fig.~16.  Comparing GX~1+4 with the data for XTE~J1743--363 in Fig.~7,
we see two things in common: a noticeable autocorrelation out to
100~dy (while even the very bright blue supergiant system 4U~1700--377
only correlates out to 30~dy) and a long quiet period (for GX~1+4, the
years 2007 and 2008 and for XTE~J1743--363 the last 5~yr).  These
common characteristics lend support to the spectroscopic evidence that
XTE~J1743--363 is also a symbiotic binary.  Strong emission lines in the
optical spectrum and pulsations of GX~1+4 have been taken as evidence
of an accretion disc \citep{chakrabarty97,jablonski97}.  While these
characteristics have not been seen in XTE~J1743--363, we note that our
far-red spectrum of XTE~J1743--363 was taken in the middle of the
extended X-ray quiet period still in progress.  It is possible that a
disc forms when the source is X-ray active.

The time variability of XTE~J1743--363 also resembles that of a member of
another class of binary entirely: the ultracompact binary 4U~0513--40
in globular cluster NGC~1851.  It also shows highly variable emission,
correlations over many days, and a relatively quiet period of approximately
three years \citep{maccarone10}. Our spectroscopy of the
counterpart rules out this possibility, unless the correct companion is
the faint, nondescript G/K star nearby.  It is unlikely, however, that a
star as unusual as the M8 giant would coincide with the X-ray source by
accident, while it is quite likely that the dwarf star would do so.

\subsection{XTE~J1901+014}

This system showed a short ($<3.15$~hr), bright (0.9~Crab) X-ray
outburst at the time of its discovery by \citet{remillard02}.  It has
since shown further outbursts \citep{krimm2010} but seems to have a
relatively high flux in quiescence as well 
\citep{karasev07,karasev08} compared to the SFXTs
\citep{smith07,krimm2010}.  It does not have an
IR/optical counterpart bright enough to be a supergiant in our Galaxy
\citep{smith07,karasev08}.  Although we didn't see an outburst in the
bulge scan data, we confirm the presence of a nonzero quiescent
baseline that is not consistent with the behaviour of the SFXTs in our
sample. \citet{karasev07}, using pointed observations of the {\it
RXTE} PCA, showed that there is a great deal of variability (by more
than a factor of two over tens of seconds) even outside of outbursts.
They quoted a typical flux of 2.7~mCrab outside of outburst.  Our
average luminosity of 5.72 $\times 10^{34}$ erg~s$^{-1}$ at an
arbitrary distance of 3~kpc corresponds to 5.3 $\times 10^{-11}$
erg~s$^{-1}$ or a bit over 2~mCrab, in agreement with
\citet{karasev07}.  The persistent emission outside of outburst
supports the more definitive evidence of the lack of a supergiant
companion in separating this source from the class of SFXTs. The
nature of the system remains a mystery.

\section*{Acknowledgments}

The authors thank Sebastian Drave for his assistance in identifying an
independent set of outbursts of XTE~J1739--302 from those used in
\citet{drave10} and for useful discussions.  DMS would also like to
thank Tom Maccarone for stimulating discussions on the interpretation
of the XTE~J1743--363 data.  We thank the referee for many perceptive
comments and for catching a number of minor errors.

The 2MASS image (Fig.~13)
was retrieved from the NASA/ IPAC Infrared Science Archive, 
which is operated by the Jet Propulsion Laboratory, California 
Institute of Technology, under contract with the National Aeronautics 
and Space Administration.

Our results on XTE~J1743--363 were partially based on observations
made with the Nordic Optical Telescope, operated on the island of La
Palma jointly by Denmark, Finland, Iceland, Norway, and Sweden, in the
Spanish Observatorio del Roque de los Muchachos of the Instituto de
Astrofisica de Canarias.  The data presented here (Figs.~14 \& 15) have been
taken using ALFOSC, which is owned by the Instituto de Astrofisica de
Andalucia (IAA) and operated at the Nordic Optical Telescope under
agreement between IAA and the Niels Bohr Institute for Astronomy,
Physics and Geophysics (NBIfAFG) of the Astronomical Observatory
of Copenhagen.
 
IN is partially supported by the Spanish Ministerio de Ciencia y
Tecnolog\'{\i}a under grants AYA2010-21697-C05-05
and CSD2006-70.

\label{lastpage}

\end{document}